\documentclass[doublecol]{epl2}

\title{Coherent and Incoherent Vortex Flow States in Crossed Channels} 
\shorttitle{Vortex Flow States in Crossed Channels}
\author{C.J. Olson Reichhardt and C. Reichhardt} 
\shortauthor{C.J.O. Reichhardt and C. Reichhardt}

\institute{ 
Theoretical Division, 
Los Alamos National Laboratory, Los Alamos, New Mexico 87545
}

\abstract{
We examine vortex flow states in periodic square pinning arrays with one
row and one column of pinning sites removed to create an easy flow
crossed channel geometry.
When a drive is simultaneously applied along both 
major symmetry axes of the pinning array
such that vortices move in both channels, 
a series of coherent flow states 
develop in the channel intersection at rational ratios of 
the drive components in each symmetry direction
when the vortices can cross the intersection without local collisions. 
The coherent flow states are correlated with
a series of anomalies in the velocity force curves, 
and in some cases can produce
negative differential conductivity. 
The same general behavior could also be realized in
other systems including
colloids, particle traffic in microfluidic devices,  
or Wigner crystals in crossed one-dimensional channels.
}

\begin{document}

\maketitle

A wide variety of different types of vortex commensurability and dynamics 
can be realized in superconductors containing a periodic array of 
artificial pinning sites\cite{Baert,Harada}. 
For fields at which the number of vortices is an integer 
multiple or rational
fraction of the number of pinning sites,
various types of vortex crystalline
states occur which are associated with peaks in the critical current 
\cite{Baert,Harada,Baert2,Field,Karapetrov,Commensurate,Peeters}. 
When there are more vortices than pinning sites,
it is possible to have highly mobile interstitial vortices 
between vortices located at the pinning
sites \cite{Harada,Karapetrov,Look,Olson}.        
One-dimensional interstitial vortex motion between pinned vortices
has been realized in systems 
with artificially fabricated weak pinning channels 
\cite{Kes,Ge,Yu,Kes2}. In these channel geometries, 
oscillations in the critical current and resistance can arise as a function
of vortex density due to changes in the number of vortex rows
moving within the channels. 

One  
aspect of vortex dynamics 
in periodic pinning arrays that
has not been studied is 
vortex motion in systems where 
groups of pins are removed to form easy flow channels 
for interstitial vortices. 
In this work we examine the vortex motion
in a system with a square pinning array 
where a row and a column of pins are removed to create
intersecting channels.
Vortices moving in the two channels 
are forced to interact at the channel intersection and can
experience interference phenomena.
Such a pinning geometry is very feasible 
to create experimentally, and the removal
of individual pinning sites 
to create diluted pinning arrays has already been achieved \cite{Koelle}.
Driving of vortices with crossed external drives and measurement of the
perpendicular voltage responses has also been demonstrated
\cite{Vicent}.       

We first study the effective matching field 
where the number of vortices equals the number of pinning sites in the
unmodified square array and there are an equal number of interstitial
vortices in each channel.
We fix an external drive along one of the channel directions such that
the vortices in that channel flow, and then slowly increase a second
drive in the perpendicular direction, causing the vortices in the
second channel to flow and interact with the moving vortices in the first
channel.
We find a series of transitions between 
disordered flows where the crossing vortices collide
and ordered or coherent flows where the
vortex motion through the intersection is synchronized and no collisions
occur.
The ordered phases create a 
series of anomalies in both velocity components. 
We also find negative differential resistance upon passing into and out of
the ordered phases
when vortices in one channel must slow down as 
the drive is increased in order to maintain an ordered flow.
The overall structure of the transport curve is a devil's staircase. 
It is distinct from the devil's staircase structures 
found for vortices \cite{Nori} 
or colloids \cite{Grier,Lacasta} moving over periodic pinning arrays.
In Refs.~\cite{Nori,Grier,Lacasta}, 
the velocity vector locks to symmetry directions of the
pinning array, so
the devil's staircase structure occurs 
even in the limit of a single isolated particle moving over the array.
In our system, the locking phases vanish in the 
limit of a single interstitial particle
moving in the channels and are replaced by
what we term a {\it vortex gating effect} 
in which the motion abruptly switches from one direction to the other.
The dynamics of vortices in structured pinning geometries can also be 
generalized to other systems including
colloids in periodic trap arrays \cite{Grier,Brunner} or
narrow channels \cite{Leid}, particle or bubble flow 
in microfluidic devices \cite{Bell}, 
and Wigner crystals in narrow channels \cite{Peeters2}. 
We expect that the results we describe here
can also be realized in 
these other systems where intersecting quasi-one-dimensional 
particle states can be constructed.

\begin{figure}
\includegraphics[width=\columnwidth]{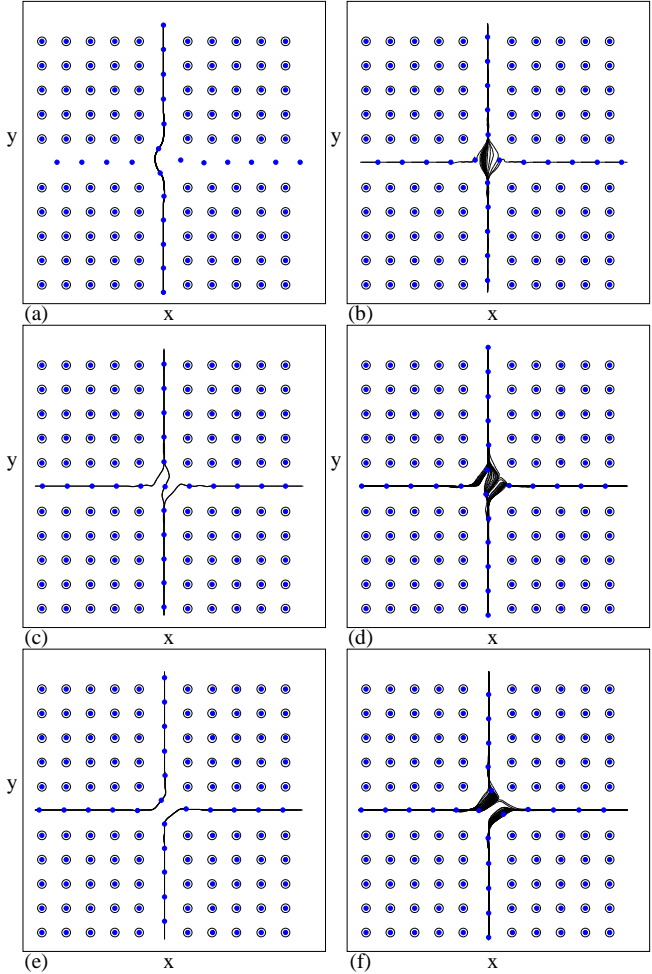}
\caption{
Pinning site locations (open circles), 
vortex locations (filled dots), and vortex trajectories (lines)
at $B=B_\phi^S$ 
for a square pinning array with one row and one column of 
pins removed. 
The vortex density matches the original square pinning
array density $B_\phi^S$.
A constant drive of $F^{D}_{y} = 0.2$ is applied in the $y$ direction and 
an increasing drive
$F^{D}_{x}$ is applied in the $x$ direction. 
(a) At $F^{D}_{x} = 0.025$,  the vortices in the $x$ channel are
pinned. 
(b) At $F^{D}_{x} = 0.04$, the 
vortex trajectories at the intersection are disordered.
(c) At $F^{D}_{x} = 0.106$ an ordered or coherent flow phase occurs. 
(d) Disordered flow at $F^{D}_{x} = 0.1105$.
(e) Ordered flow at $F^{D}_{x} = 0.2$.
(f) Disordered flow at $F^{D}_{x} = 0.224$.
}
\label{fig_schematics}
\end{figure}

{\it Simulation-} We consider a two-dimensional system 
with periodic boundary conditions in the
$x$ and $y$ directions. The sample geometry is illustrated 
in fig.~1(a).  A single column and a single row of pinning sites are removed
from a square pinning array to create an intersection at the center
of the sample.
The number of vortices $N_v$ is equal to the number of pinning sites that would
have been present in the original
square pinning array, and is larger than the actual 
number of pinning sites $N_p$ in the modified square array.
That is, the applied field $B=B_\phi^S$, where $B_\phi^S$ is the
matching field of the undiluted square array.
All of the 
pinning sites are occupied and 
the remaining  vortices are located in the interstitial regions
in the empty row and column. 
A single vortex $i$ located at position ${\bf R}_i$ 
obeys the  overdamped equation of motion:
\begin{equation}
\eta\frac{ d{\bf R}_{i}}{dt} = {\bf F}_{i}^{vv}  
+ {\bf F}^{p}_{i} + {\bf F}^{D}_{i} .
\end{equation}
The repulsive vortex-vortex interaction force is given by
${\bf F}_{i}^{vv} = \sum^{N_{v}}_{j\ne i}A 
K_{1}(R_{ij}/\lambda){\bf {\hat R}}_{ij},$
where $K_{1}$ is the modified Bessel function, 
$A=\phi_0^2/(2\pi\mu_0\lambda^3)$, 
$\phi_0=h/2e$ is the flux quantum,
 $\lambda$ is the London penetration depth, 
$R_{ij}=|{\bf R}_i-{\bf R}_j|$,
and ${\bf {\hat R}}_{ij}=({\bf R}_i-{\bf R}_j )/R_{ij}$.
The damping constant $\eta=\phi_0^2d/2\pi\xi^2\rho_N$,
where
$\xi$ is the
superconducting coherence length, $\rho_N$ is the normal state
resistivity of the material, and $d$ is the thickness of the superconducting
crystal.
The pinning force ${\bf F}_i^{p}$ arises from attractive parabolic potential
wells arranged as shown in fig.~1 with a maximum force of $F_p$ and a
radius of $R_p=0.35\lambda$,
${\bf F}_i^{p}=\sum_{k=1}^{N_p}AF_p(R_{ik}^{p}/R_p)\Theta((R_p-R_{ik}^p)/\lambda)
{\bf {\hat R}}_{ik}^p$.
Here $\Theta$ is the Heaviside step function, ${\bf R}_k^p$ is the location of
pinning site $k$, $R_{ik}^p=|{\bf R}_i-{\bf R}_k^p|$, and
${\bf {\hat R}}_{ik}^p=({\bf R}_i-{\bf R}_k^p)/R_{ik}^p$.
The external force 
${\bf F}^D_{i} = A(F^{D}_{x}{\hat {\bf x}} + F^{D}_{y}{\hat {\bf y}})$
represents 
the Lorentz force from an applied current. 
The vortex velocities 
$V_x=(N_v)^{-1}\sum_{i=1}^{N_v}{\hat {\bf x}} \cdot d{\bf R}_i/dt $ and
$V_y=(N_v)^{-1}\sum_{i=1}^{N_v}{\hat {\bf y}} \cdot d{\bf R}_i/dt $
in the two directions represent
the resulting voltages. 
In this work, all driving forces are small enough 
that the vortices at the pinning sites remain pinned.
The initial vortex positions are found by simulated annealing 
in a procedure similar to that used previously \cite{Olson}.
After the vortex positions are initialized,
we apply a fixed force $F_y^{D}$ in the positive $y$-direction 
while slowly increasing a force $F_x^{D}$ in the 
positive $x$ direction.

\begin{figure}
\includegraphics[width=\columnwidth]{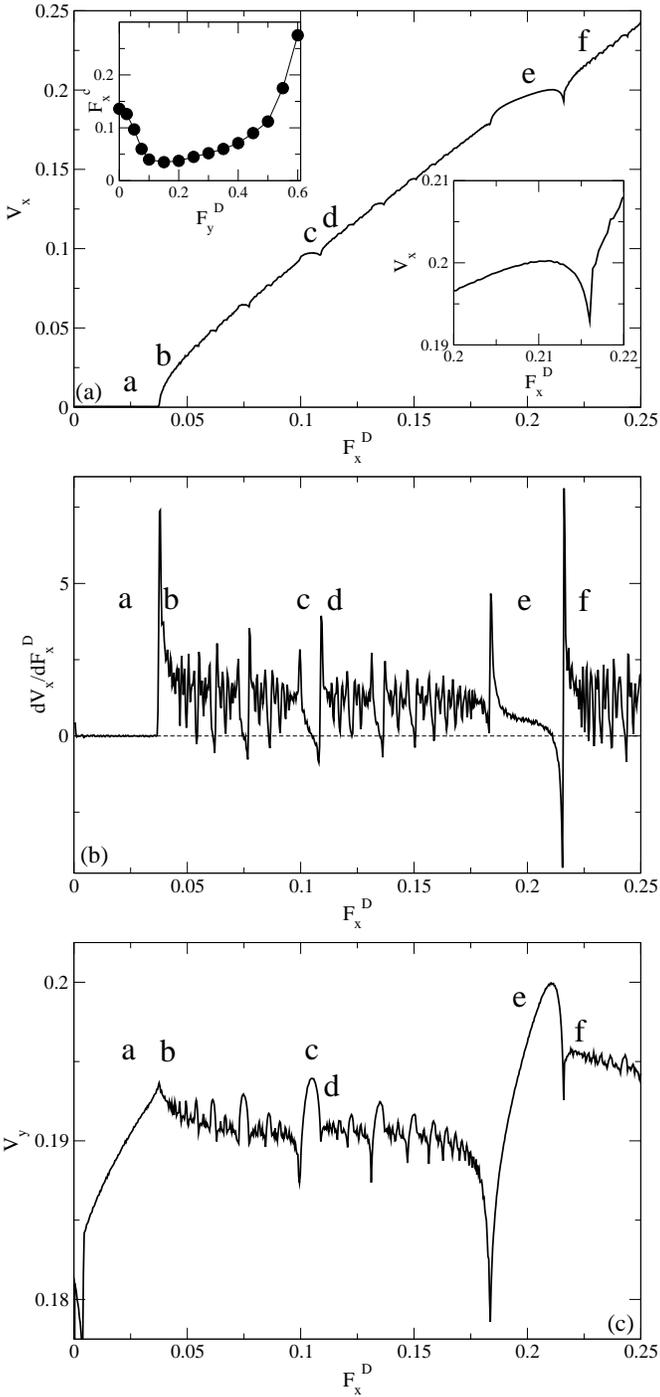}
\caption{(a) $V_{x}$ vs $F^{D}_{x}$, (b) $dV_{x}/dF^{D}_{x}$ vs $F^D_x$, 
and (c) $V_{y}$ vs $F^{D}_{x}$ for the
system in fig.~1 with $F_y^D=0.2$. 
The points marked {\bf a}-{\bf e} correspond to 
the flow regimes illustrated in fig.~1(a-e). 
Upper left inset in (a): 
depinning threshold $F_x^c$ for motion in the $x$-direction
for varied values of $F^{D}_{y}$ for the same system.
Lower right inset in (a): a blowup
of $V_{x}$ vs $F^{D}_{x}$ from the main panel 
highlighting the decrease in $V_{x}$ with increasing $F^{D}_{x}$ as the
system exits the $F^{D}_{y}/F^{D}_{x} = 1.0$ locked phase. 
In (b) several regions of negative
differential conductivity appear with $dV_{x}/dF^{D}_{x}  < 0$.   
}
\label{fig_grainboundaries}
\end{figure}

We first consider the vortices under matching conditions
$B=B_\phi^S$, where the number
of vortices equals the number of pinning sites in the original square pinning
array.
In the absence of a drive, the vortices form a square lattice with the
interstitial vortices sitting at the locations of the pinning sites that
were removed.
We illustrate the vortex trajectories for several values of $F_x^D$
at fixed $F_y^D=0.2$ in fig.~1,
and we plot 
$V_{x}$ versus $F^{D}_{x}$ in fig.~2(a), 
$dV_{x}/dF^{D}_{x}$ versus $F^{D}_{x}$ in fig.~2(b), and 
$V_{y}$ versus  $F^{D}_{x}$ in fig.~2(c)
for the same system.
The points marked {\bf a}-{\bf e} in fig.~2 correspond to the 
values of $F^{D}_{x}$ illustrated in 
fig.~1(a-e). 

At $F^{D}_{x} = 0$, the vortices 
in the $y$ channel are flowing while 
the vortices in the $x$ channel are pinned. 
The vortices in the $x$ channel do not depin until $F_x^D$ reaches a
critical value
$F^{c}_{x}=0.037$.
For $F^{D}_{x} < 0.037$ the vortices
in the $x$ channel are pinned but those in the $y$ channel are moving, as 
indicated by the fact that $V_{x} = 0$ over this regime in fig.~2(a) while 
$V_{y}$ has a finite value in fig.~2(c). 
Figure~1(a) illustrates
the vortex trajectories 
for $F^{D}_{x} = 0.025$ where the flow is confined to the $y$ direction.
At $F^D_x=0.4$, fig.~1(b) shows that
vortices in both channels
are moving and that the trajectories are disordered at the 
channel intersection. 

In fig.~1(c) at $F^{D}_{x}  = 0.106$,
the vortex flow at the intersection is ordered or
coherent and consists of a mixing flow 
where vortices approaching the intersection from the left turn upwards into
the $y$ channel, while a portion of the
vortices approaching the intersection from below turn
right into the $x$ channel.
The net flow at the intersection 
consists of two vortices exiting in the positive $y$ direction 
for every one vortex exiting in the positive $x$ direction. 
The ordered flow regime correlates with a steplike transport anomaly
centered around $F^{D}_{x} = 0.1$ at the point marked {\bf c} in fig.~2(a). 
Two peaks in $dV_x/dF_x^D$ in fig.~2(b) surround this
region, while a smooth rounded peak appears in $V_{y}$
in fig.~2(c). This shows that the vortices remain locked in the
ordered flow phase illustrated in fig.~1(c) 
for a range of $F^{D}_{x}$. 
In general, ordered flow phases are associated with all of
the transport anomalies in fig.~2. The phases are centered
at rational ratios of the external drives 
$F^{D}_{y}/F^{D}_{x} = n/m$ with $n,m$ integer, 
such as at $F^D_y/F^D_x=1/4$, $1/2$, or $1.0$.  
Figure~1(c) shows the
$F^D_y/F^D_x=1/2$ state. 

In fig.~2, anomalies at other rational ratios of $F^D_y/F^D_x$
are visible, with the most
pronounced steplike features occurring for cases where 
$n<4$ and $m<4$.
This type of behavior is very similar to the
mode locking found in systems which contain
an intrinsic oscillating frequency that locks
to an external ac drive \cite{Kes2}. 
Another intriguing feature 
is the fact that near the drive at which the system exits a
locked phase, such as near the ends of the $F^D_y/F^D_x=1/2$ and $1.0$
steps, 
$V_x$ {\it decreases} with increasing $F^{D}_{x}$. 
This is 
highlighted in the lower right inset of 
fig.~2(a) for the end of the $F^D_y/F^D_x=1.0$ locked phase. 
The velocity decrease
produces {\it negative differential conductivity} 
where $dV_{x}/dF^{D}_{x} < 0$, as shown in fig.~2(b). 
This indicates that 
in order for the system to remain locked as it approaches
the upper end of the locked phase, 
the vortices must move more slowly in the $x$ direction
as $F^D_x$ increases.         

The anomalies we find in the velocity force curves 
are similar to those observed for vortices
or colloids driven over periodic pinning arrays 
when the particle velocity vector
locks to symmetry directions of the
pinning array \cite{Nori,Grier,Lacasta}. 
In these systems, the locking produces clear steps and occurs even in 
the limit of an isolated driven particle \cite{Lacasta}. 
The phase locking considered here
for the crossed channel geometry results from a different 
mechanism and does not occur in the
single particle 
limit, as we describe below in fig.~3. 
In addition, there was no negative differential conductivity associated with
the phase locking in 
refs.~\cite{Nori,Grier,Lacasta}. 

Away from the steplike velocity anomalies, 
the vortex trajectories are disordered, as shown in
fig.~1(d) for $F^{D}_{x} = 0.1105$. 
Figure~1(e) illustrates the ordered flow phase at 
$F^{D}_{x} = 0.2$, which corresponds to the 
pronounced $F^{D}_{y}/F^{D}_{x} = 1.0$ anomaly in fig.~2. 
Also associated with this ordered flow is the dip in $V_x$ highlighted in
the lower inset of fig.~2(a), the arctangent-like feature 
and large negative differential conductivity in $dV_x/dF_x^D$
in fig.~2(b), and the smooth feature in $V_y$ at point {\bf e} in fig.~2(c).
Figure~1(f) shows the 
disordered vortex trajectories at $F^{D}_{x} = 0.224$, just outside
of the $F^{D}_y/F^D_x=1.0$ state.

\begin{figure}
\includegraphics[width=\columnwidth]{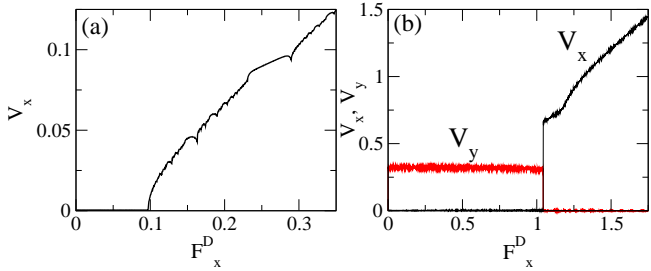}
\caption{(a) $V_{x}$ vs $F^{D}_{x}$ for the same system in 
fig.~2(a) but with $F^{D}_{y} = 0.1$. 
The initial pinned phase 
and the $F^{D}_{y}/F^{D}_{x} = 1.0$ state are enhanced. 
(b) $V_{x}$ (dark line) and $V_{y}$ (light line) 
vs $F^{D}_{x}$ at $F^D_y=0.7$ with only a
single interstitial vortex in the channels. 
All the anomalies seen in fig.~2 are lost and there is only a sharp transition
from $y$ direction motion to 
$x$ direction motion as a function of $F^{D}_{x}$.
}
\label{fig_plots}
\end{figure}

The general features we find in the transport 
characteristics are independent of the rate at which $F^{D}_{x}$ is increased,
and we find similar features if we vary $F^D_y$ at fixed $F^D_x$. 
In the upper left 
inset of fig.~2(a) we show the 
depinning force $F_x^c$ for motion in the $x$ channel vs $F^{D}_{y}$. 
As $F^D_y$ increases from zero, $F_x^c$ initially decreases and then begins
to increase.
The increase occurs when the $y$ direction driving force
shifts the vortices in the $x$ channel 
closer to the occupied pinning sites along the upper side of the channel, 
increasing the coupling between the
interstitial and pinned vortices and thus increasing the
depinning force for the interstitial vortices.
The widths of the velocity anomalies are enhanced 
when the depinning threshold increases.  This is illustrated
in fig.~3(a) where we plot $V_{x}$ versus $F^{D}_{x}$ for the
same system in fig.~2(a) but with $F^{D}_{y} = 0.1$.
Both the pinned phase and the $F_y^D/F_x^D=1.0$ state are
broader
at $F^D_y=0.1$ than at $F^D_y=0.2$.

We next examine the case of a single interstitial vortex driven 
through the channels. We remove all but one of the interstitial vortices
and initialize this single interstitial vortex in the $y$ channel.
In fig.~3(b) we plot $V_{x}$ and $V_{y}$ versus 
$F^{D}_{x}$ at $F^{D}_{y} = 0.7$ 
showing that in the single particle limit, 
all the transport anomalies seen in fig.~2 are lost and there is simply 
a transition from pure $y$ direction motion to pure $x$ direction
motion 
when $F^{D}_{x}$ increases.
This is indicated by the sharp transition to $V_{y} = 0.0$ 
which coincides with the onset of a finite value of  $V_{x}$. 
We term this sharp transition
a vortex gating effect. 

\begin{figure}
\includegraphics[width=\columnwidth]{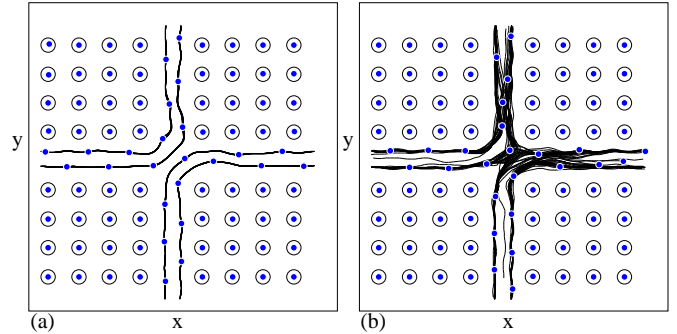}
\caption{
Vortex positions (filled dots), 
pinning site locations (open circles), and trajectories (lines) 
for the same system in fig.~1 but with a higher vortex 
density of $B/B_{\phi}^S = 1.067$. Here the vortices in the
channel undergo a buckling transition to a zig-zag pattern. 
(a) Ordered flow at the $F^{D}_{y}/F^{D}_{x} = 1.0$ state.     
(b) Disordered flow at  $F^{D}_{y}/F^{D}_{x} = 0.92$.
 }
\label{fig_four}
\end{figure}

The loss of the phase locking for a single interstitial vortex indicates that
collective interactions are responsible for the phase locking in the crossed
channel geometry.
The discreetness of the vortices arriving at the intersection of the channels
permits only an integer number of vortices to move in each of 
the two directions. 
The ordered trajectories such as those seen in fig.~1(c,e) 
distort somewhat but stay ordered as $F^{D}_{x}$ 
increases, allowing the interstitial vortices to remain 
in the same phase-locked state over a
wider range of $F^{D}_{x}$. 
The distortion of the vortex trajectories is also responsible for the
negative differential conductivity 
and the changes in $V_{y}$ associated with the coherent flow phases. 
In the case of vortices or colloids moving over periodic pinning sites
in Refs.~\cite{Nori,Lacasta},
the ordered flow phases consisted of one-dimensional 
paths oriented along different pinning lattice symmetry angles. 
These paths cannot distort and so there   
was no differential negative conductivity in these systems.   

The phase locking illustrated in figs. 1 and 2 
occurs for a range of fields and is most prominent for
$0.925 < B/B_{\phi}^S < 1.1$. 
Within this field range, the interstitial vortices are confined in the
channels. 
In fig.~4(a,b) we plot the vortex trajectories for 
a sample with $B/B_{\phi}^S =1.067$. 
In this case, the vortices
in the channel undergo a {\it buckling} transition and 
form a zig-zag pattern. 
fig.~4(a) illustrates the $F^{D}_{y}/F^{D}_{x} = 1.0$ state  
where an ordered flow pattern occurs similar to that seen in
fig.~1(e). 
Here there are two staggered moving rows of vortices in each channel.
For $F^{D}_{y}/F^{D}_{x} =0.92$, shown in fig.~4(b), 
the vortex trajectories are 
disordered and there is no steplike anomaly in the transport curves at
this point.
For $B/B_{\phi}^S > 1.1$, some interstitial vortices begin to penetrate the bulk 
pinned region away from the channels. 
These interstitial vortices are generally more strongly pinned 
than the vortices in the channels; however,
for intermediate ranges of $F^{D}_{x}$ and $F^{D}_{y}$ 
they can become mobile and add extra features to the 
velocity-force curves. These effects are beyond the scope of this work.

In summary, 
we studied a square periodic pinning array with crossed easy flow
channels for vortex motion formed 
by the removal of one row and one column of pinning.
We show that when a fixed drive is applied in one direction and 
a slowly increasing drive is applied 
in the perpendicular direction, a 
series of phases with coherent vortex flow dynamics occurs 
which are centered at rational ratios of the two drives. 
These coherent phases appear as anomalies in the
transport curves and are also associated with 
negative differential conductivity. The
phase locking anomalies are distinct from those 
observed for the velocity vector locking of particles driven 
at different angles over 
periodic pinning arrays,
which appear even in the single particle case. 
For the crossed channel geometry, the
phase locking occurs due to the discrete nature of the particles. 
In the single particle limit, the transport anomalies vanish 
and are replaced by a vortex gating phenomena
in which the motion switches abruptly from one channel to the other. 
We expect that similar effects can occur 
for particle transport in other systems with
quasi-one-dimensional crossed geometries, 
such as colloidal systems, microfluids, or crossed 
one-dimensional  Wigner crystal systems. 

\acknowledgments
This work was carried out under the auspices of the NNSA of the U.S. DoE
at LANL under Contract No. DE-AC52-06NA25396.


\begin{thebibliography}{99}

\bibitem{Baert}
  \Name{Baert M., Metlushko V.V., Jonckheere R., Moshchalkov V.V. \and
Bruynseraede Y.}
  \REVIEW{Phys. Rev. Lett.}{74}{1995}{3269};
  \Name{Mart{\' \i}n J.I., V{\' e}lez M., Nogu{\' e}s J. \and 
Schuller I.K.}
  \REVIEW{Phys. Rev. Lett.}{79}{1997}{1929};
  \Name{Metlushko V.V., DeLong L.E., Baert M., Rosseel E., Van Bael M.J.,
Temst K., Moshchalkov V.V. \and Bruynseraede Y.}
  \REVIEW{Europhys. Lett.}{41}{1998}{333};
  \Name{Welp U., Xiao Z.L., Jiang J.S., Vlasko-Vlasov V.K., Bader S.D.,
Crabtree G.W., Liang J., Chik H. \and Xu J.M.}
  \REVIEW{Phys. Rev. B}{66}{2002}{212507}.

\bibitem{Harada}
  \Name{Harada K., Kamimura O., Kasai H., Matsuda T., Tonomura A. \and
Moshchalkov V.V.}
  \REVIEW{Science}{274}{1996}{1167}.

\bibitem{Baert2}
  \Name{Baert M., Metlushko V.V., Jonckheere R., Moshchalkov V.V. \and
Bruynseraede Y.}
  \REVIEW{Europhys. Lett.}{29}{1995}{157}.

\bibitem{Field}
  \Name{Field S.B., James S.S., Barentine J., Metlushko V., Crabtree G.,
Shtrikman H., Ilic B. \and Brueck S.R.J.}
  \REVIEW{Phys. Rev. Lett.}{88}{2002}{067003};
  \Name{Grigorenko A.N., Bending S.J., Van Bael M.J., Lange M., 
Moshchalkov V.V., Fangohr H. \and de Groot P.A.J.}
  \REVIEW{Phys. Rev. Lett.}{90}{2003}{237001}.

\bibitem{Karapetrov} 
  \Name{Karapetrov G., Fedor J., Iavarone M., Rosenmann D. \and Kwok W.K.}
  \REVIEW{Phys. Rev. Lett.}{95}{2005}{167002}.

\bibitem{Commensurate}
  \Name{Reichhardt C., Olson C.J. \and Nori F.}
  \REVIEW{Phys. Rev. B}{57}{1998}{7937};
  \Name{Reichhardt C. \and Gr{\o}nbech-Jensen N.}
  \REVIEW{Phys. Rev. B}{63}{2001}{054510}.

\bibitem{Peeters}
  \Name{Berdiyorov G.R., Milosevic M.V. \and Peeters F.M.}
  \REVIEW{Phys. Rev. Lett.}{96}{2006}{207001}.

\bibitem{Look}
  \Name{Rosseel E., Van Bael M., Baert M., Jonckheere R., Moshchalkov V.V.
\and Bruynseraede Y.}
  \REVIEW{Phys. Rev. B}{53}{1996}{R2983};
  \Name{Van Look L., Rosseel E., Van Bael M.J., Temst K., Moshchalkov V.V.
\and Bruynseraede Y.}
  \REVIEW{Phys. Rev. B}{60}{1999}{R6998};
  \Name{Silhanek A.V. \it{et al.}}
  \Book{{\rm to be published}}.

\bibitem{Olson}
  \Name{Reichhardt C., Olson C.J. \and Nori F.}
  \REVIEW{Phys. Rev. Lett.}{78}{1997}{2648};
  \Name{Misko V.R., Savel'ev S., Rakhmanov A.L. \and Nori F.}
  \REVIEW{Phys. Rev. Lett.}{96}{2006}{127003};
  \Name{Reichhardt C. \and Olson Reichhardt C.J.}
  \REVIEW{Phys. Rev. B}{79}{2009}{134501}.

\bibitem{Kes} 
  \Name{Besseling R., Kes P.H., Dr{\" o}se T. \and Vinokur V.M.}
  \REVIEW{New J. Phys.}{7}{2005}{71}.

\bibitem{Ge} 
  \Name{Grigorieva I.V., Geim A.K., Dubonos S.V., Novoselov K.S., 
Vodolazov D.Y., Peeters F.M., Kes P.H. \and Hesselberth M.}
  \REVIEW{Phys. Rev. Lett.}{92}{2004}{237001}.

\bibitem{Yu}
  \Name{Yu K., Heitmann T.W., Song C., DeFeo M.P., Plourde B.L.T.,
Hesselberth M.B.S. \and Kes P.H.}
  \REVIEW{Phys. Rev. B}{76}{2007}{220507}.

\bibitem{Kes2}
  \Name{Kokubo N., Besseling R., Vinokur V.M. \and Kes P.H.}
  \REVIEW{Phys. Rev. Lett.}{88}{2002}{247004};
  \Name{Kokubo N., Sorop T.G., Besseling R. \and Kes P.H.}
  \REVIEW{Phys. Rev. B}{73}{2006}{224514}.

\bibitem{Koelle}
  \Name{Kemmler M., Bothner D., Ilin K., Siegel M., Kleiner R. \and Koelle D.}
  \REVIEW{Phys. Rev. B}{79}{2009}{184509}.

\bibitem{Vicent}
  \Name{Silhanek A.V., Van Look L., Raedts S., Jonckheere R. \and
Moshchalkov V.V.}
  \REVIEW{Phys. Rev. B}{68}{2003}{214504};
  \Name{Villegas J.E., Gonzalez E.M., Montero M.I., Schuller I.K. \and
Vicent J.L.}
  \REVIEW{Phys. Rev. B}{72}{2005}{064507};
  \Name{Gonzalez E.M., Nunez N.O., Anguita J.V. \and Vicent J.L.}
  \REVIEW{Appl. Phys. Lett.}{91}{2007}{062505}.

\bibitem{Nori} 
  \Name{Reichhardt C. \and Nori F.}
  \REVIEW{Phys.~Rev.~Lett}{82}{1999}{414}.

\bibitem{Grier}
  \Name{Korda P., Taylor M.B. \and Grier D.G.}
  \REVIEW{Phys. Rev. Lett.}{89}{2002}{128301};
  \Name{MacDonald M.P., Spalding G.C. \and Dholakia K.}
  \REVIEW{Nature (London)}{426}{2003}{321};
  \Name{Gopinathan A. \and Grier D.G.}
  \REVIEW{Phys. Rev. Lett.}{92}{2004}{130602}.

\bibitem{Lacasta}
  \Name{Lacasta A.M., Sancho J.M., Romero A.H. \and Lindenberg K.}
  \REVIEW{Phys. Rev. Lett.}{94}{2005}{160601};
  \Name{Hermann J., Karweit M. \and Drazer G.}
  \Book{\rm {arXiv:0904.2538 preprint (2009)}}.

\bibitem{Brunner}
  \Name{Mangold K., Leiderer P. \and Bechinger C.}
  \REVIEW{Phys. Rev. Lett.}{90}{2003}{158302}.

\bibitem{Leid}
  \Name{K{\" o}ppl M., Henseler P., Erbe A., Nielaba P. \and Leiderer P.}
  \REVIEW{Phys. Rev. Lett.}{97}{2006}{208302}.

\bibitem{Bell}
  \Name{Belloul M., Engl W., Colin A., Panizza P. \and Ajdari A.}
  \REVIEW{Phys. Rev. Lett.}{102}{2009}{194502}.

\bibitem{Peeters2}
  \Name{Piacente G. \and Peeters F.M.}
  \REVIEW{Phys. Rev. B}{72}{2005}{205208}.


\end{thebibliography}
\end{document}